# Speckle Patterns and 2-Dimensional Spatial Models

J. R. Smith[a], J. J. Llovera-Gonzalez[b], S. P. Smith[c]

a) University of California at Davis, Davis, California, USA 95616.
smith@physics.ucdavis.edu †
b) Physics Department, Faculty of Electrical Engineering, Instituto Superior Politécnico "José Antonio Echeverria"
llovera@electrica.cujae.edu.cu
c) Division of Mathematics and Science Holy Names University, Oakland, California, USA 94619
hucklebird@aol.com
†correspondence author

**Abstract.** The result of 2-dimensional Gaussian lattice fit to a speckle intensity pattern based on a linear model that includes nearest-neighbor interactions is presented. We also include a Monte Carlo simulation of the same spatial speckle pattern that takes the nearest-neighbor interactions into account. These nearest-neighbor interactions lead to a spatial variance structure on the lattice. The resulting spatial pattern fluctuates in value from point to point in a manner characteristic of a stationary stochastic process. The value at a lattice point in the simulation is interpreted as an intensity level and the difference in values in neighboring cells produces a fluctuating intensity pattern on the lattice. Changing the size of the mesh changes the relative size of the speckles. Increasing the mesh size tends to average out the intensity in the direction of the mean of the stationary process.

**Sumario.**
Se presenta el resultado del ajuste de una red gaussiana bidimensional a un patrón de intensidades de speckle sobre la base de un modelo lineal de interacción de entorno cercano. Se incluye una simulación de Monte Carlo del mismo patrón de speckle que tomando en cuenta la interacción de corto alcance conduce a una estructura de varianza espacial de la red. El patrón de intensidad espacial resultante fluctúa en valores punto a punto de la manera característica de un proceso estocástico estacionario. El valor de cada punto de la red es interpretado como un nivel de intensidad. Las diferencias en los valores en las celdas colindantes producen un patrón de intensidad fluctuante sobre la red. Cambiando el tamaño de la malla cambian los tamaños relativos de los speckles. El incremento en el tamaño de la malla tiende a promediar la intensidad hacia la media del proceso estacionario.

## 1 Brownian motion models of spatial speckle patterns in time.

It is known speckle patterns are fine-granular pattern of fluctuation of intensity reflected in a surface resulting of the superposition of coherent light like a laser (see Ref. 1). Fig. 1 shows an example of speckle pattern obtained illuminating a surface of a volatile liq-uid deposited in a solid plate. The fluctuations of the spatial pattern are related to inhomogeneity in the scattering centers within the material. The spatial intensity distribution of the speckle pattern itself at any moment in time can be characterized by the methods of spatial modeling with point-to-point correlations. Such spatial modeling is the main objective of this paper.

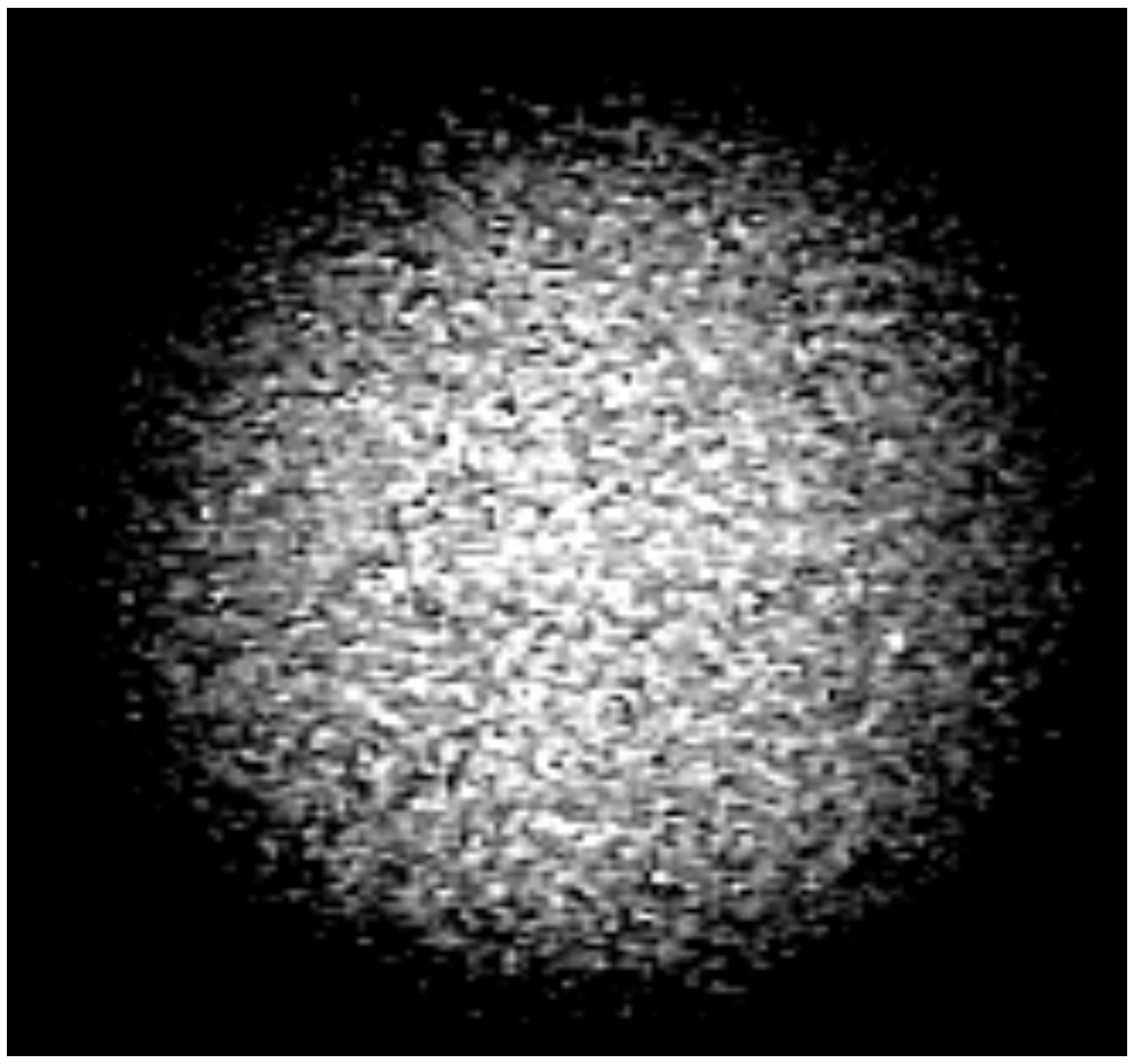

**Fig. 1.** Speckle pattern obtained illuminating with a laser beam a surface of a volatile liquid deposited in a solid plate.

The idea of characterizing the time dependence of speckle patterns by the method of Brownian motion theory has been proposed several years ago in Refs. [2] and [3]. Péron *et al*. make use of the fact that the speckle pattern contains information about the scattering/diffusion medium because the statistical properties of the speckles are related to the optical properties of the medium. Péron *et al*. employ the fractal method to approximate the diffusion process.

A stochastic process $\{X_t, t \ge 0\}$ is called a Brownian motion process if:

(i) $\{X_t, t \ge 0\}$ has stationary independent increments

(ii) For every $t > 0$, $X_t$ is normally distributed

(iii) For all, $E[X_t] = 0$

(iv) $X_0 = 0$

In the model of Péron *et al*. the effects of autocorrelation lead to diffusion and are introduced using the fractional Brownian motion model

$$E[(X_{t2} - X_{t1})^2] \propto |t_2 - t_1|^{2H}$$

where H is the Hurst exponent characterizing the autocorrelation in time. For 0 < H < 0.5 the correlations are negative, for H = 0.5 there is no correlation, and for 0.5 < H < 1 the correlations are positive.

Diffusion over time is characterized by the following diffusion function

$$D_F(\Delta t) = E\left[\left(X_{(t+\Delta t)} - X_t\right)^2\right],$$

Ref. [3] uses Monte Carlo simulations of the scattering and interfering photon packets to characterize speckle patterns. The results are summarized in terms of the speckle contrast given by

$$K^2(T) = \frac{\langle I^2 \rangle}{I^2} - 1 = \frac{2}{T}\int_0^T \beta |g_1(\tau)|^2 \left(1 - \tau/T\right) d\tau,$$

where T is the integration time and $\beta$ is the coherence factor. The light field is represented with a dynamic part and a static part and the function $g_1(\tau)$ is then split into two terms, which characterize these partial contributions. The function $g_1(\tau)$ is well represented by the stretched-exponential form derived, in the case of colloidal suspensions, from diffusing-wave spectroscopy (DWS) as

$$g_1(\tau) = \exp(-\gamma\sqrt{6\tau/\tau_0})$$

where $\gamma$ is a constant near 2 and $\tau_0$ is the relaxation time characteristic of Brownian motion in the suspension.

Péron *et al.* parameterize the speckle pattern in terms of the rows and columns of a matrix laid out in a linear array. They then substitute the spatial intensity pattern for the temporal pattern by examining this linear array. They interpret the parameters in the diffusion function in terms of the spatial correlations via the Fokker-Planck parameterization:

$$D_F(\Delta x) = G_X(1 - \exp(-\lambda_X |\Delta x|^{2\nu})$$

where $\nu$ representing a spatial analog of the Hurst parameter.

## 2. Spatial Speckle Characterization.

We apply a similar analysis to the speckle data to describe a statistical method that incorporates nearest neighbor effects in the spatial pattern. Ref. [18] points out that it is not sufficient to assume that each sample point in a speckle is independent of its neighbors in SAR Speckle Simulation (see Ref. [18]).

Fig. 2 illustrates the cell-to-cell variation in the speckle intensity pattern. The spatial correlations are visible to the eye. It is the objective of this paper to develop a method of incorporating the cell-to-cell correlations of the intensity pattern in a 2-dimensional Gaussian lattice model.

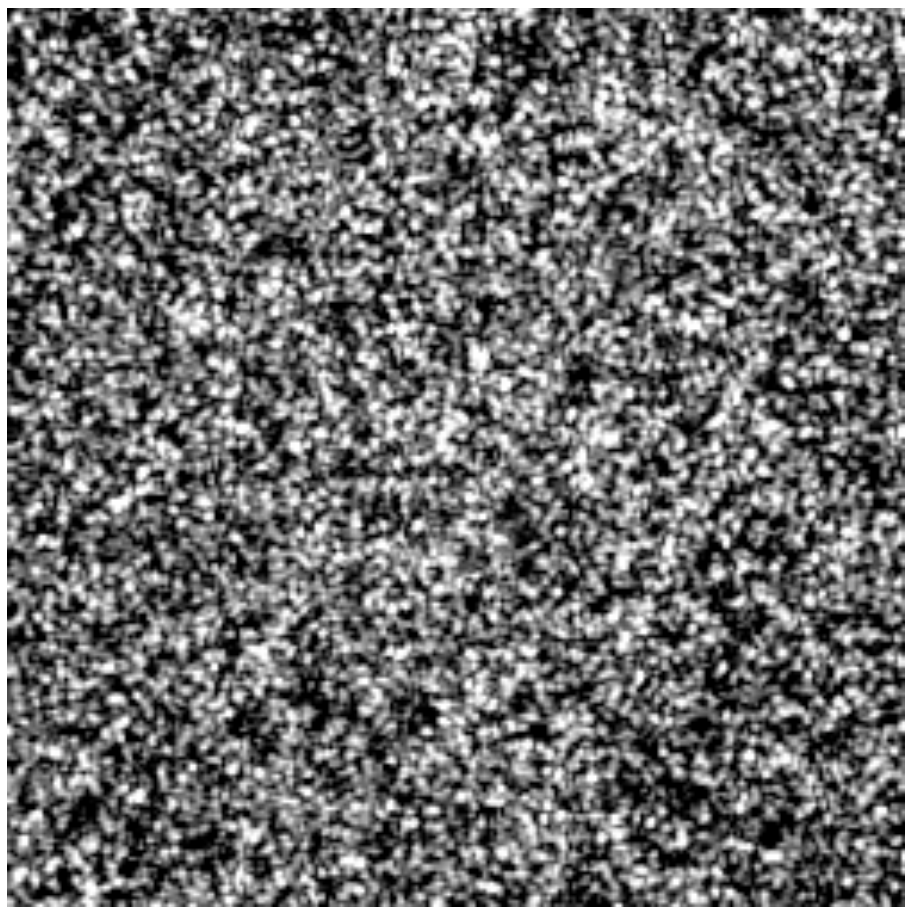

**Fig. 2.** High contrast speckle intensity pattern.

The intensity pattern corresponding to the data of Fig. 1 is plotted in Fig. 3 (in arbitrary units). The distribution is well fit to the Rician shape with parameters $I_0 = 32.397$ and $I_0 / I_{ave} = 1.867$ (see Eq. (13) in Ref. [19]).

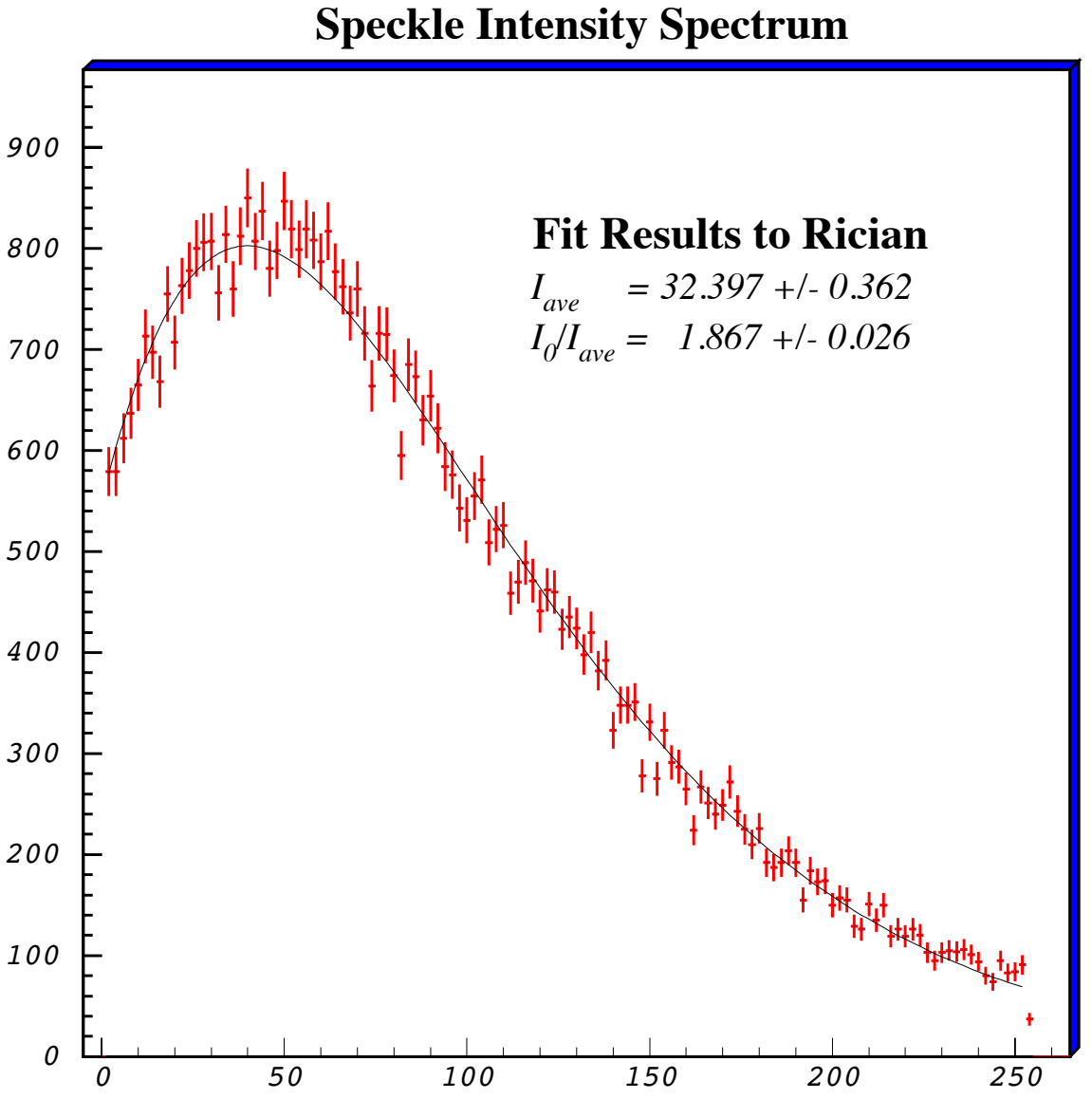

**Fig. 3.** Speckle intensity pattern obtained from the data shown in Fig. 1 with fit to a Rician distribution superimposed.

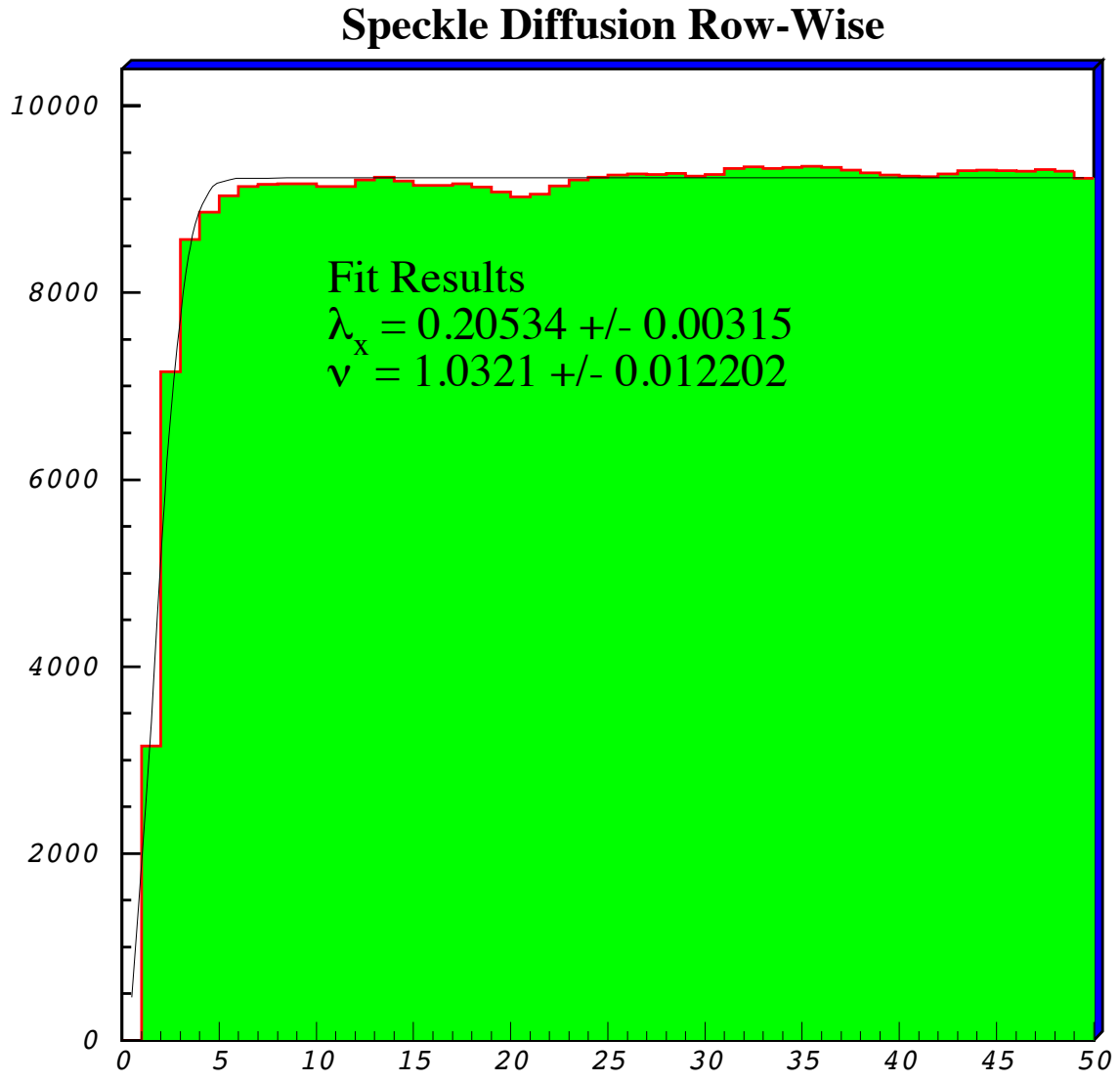

**Fig. 4.** Spatial analog of spatial variance: $D_F(\Delta x)$.

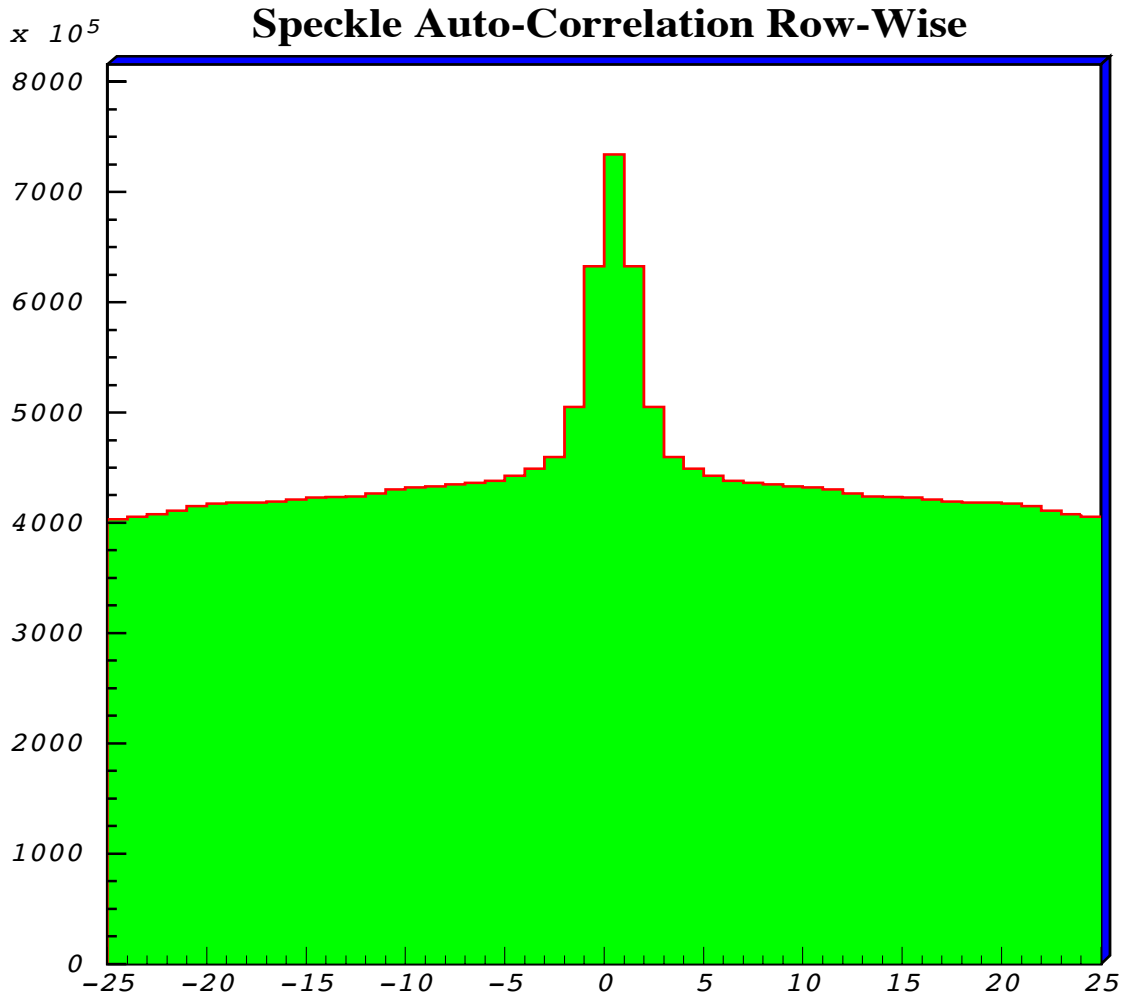

**Fig. 5.** Row-wise spatial auto-correlation in units of pixel size.

Fig. 4 shows the spatial analog of the temporal Brownian motion analysis (based on the Fokker-Planck parameterization) and indicates that there is a positive spatial correlation when the cells of the intensity pattern are laid out linearly row-by-row (spatial Hurst parameter $\nu$ is close to 1). We therefore expect that nearest neighbor effects cannot be ignored. As can be seen in Fig. 5, there is also a visible auto-correlation in the nearest cells in the intensity pattern.

We now turn to a spatial analysis based on a full lattice model including nearest neighbor lattice effects.

## 2. Mixed linear models and lattices.

In the following, we denote the transpose of a matrix $\mathbf{R}$ by $\mathbf{R}'$. We also denote column vector in component form by square brackets, $\mathbf{v} = [a,b,c,...]$, and the corresponding row vectors with parentheses, $\mathbf{v}' = (a,b,c,...)$.

A spatial variable on a rectangular lattice is represented by $u_{ij}$ which gives the value of the variable $\mathbf{u}$ at the $i-th$ row and $j-th$ column. The effects of neighbors on the value for $\mathbf{u}$ at $i, j$ can be summarized in the following form.

$$u_{ij} = \sum_{rs \in \Omega} \alpha_{rs} u_{i-r,j-s} + \sigma \varepsilon_{ij} \qquad (1)$$

where the $\varepsilon_{ij}$ are independent N(0,1) normal deviates and $\Omega$ is a finite index set for r and s which limits the extent of the influences from distant parts of the lattice. Eq. (1) can be recast into matrix form

$$\mathbf{u} = \Delta \mathbf{u} + \varepsilon \qquad (2)$$

where $E[\varepsilon] = 0$ and $\mathrm{var}[\varepsilon] = \sigma^2 \mathbf{I}$.
The form of Eq. (2) implies that the variance of u is given by

$$\mathrm{var}[\mathbf{u}] = \sigma^2 (\mathbf{I} - \Delta)^{-1} [(\mathbf{I} - \Delta)^{-1}]'$$

or that the inverse matrix is

$$\mathbf{G}^{-1} = \sigma^{-2} (\mathbf{I} - \Delta)(\mathbf{I} - \Delta)' \qquad (3)$$

Linear models, including mixed linear models, have found widespread use in statistical investigations. The linear model, though additive, is frequently flexible enough for real situations as an approximation around the mean. The mixed linear model is represented by:

$$\mathbf{y} = \mathbf{X}\beta + \mathbf{Z}\mathbf{u} + \varepsilon_{\mathbf{r}} \qquad (4)$$

Where $\mathbf{y}$ is a vector of observations, $\beta$ is a vector of fixed effects, $\mathbf{u}$ is vector of random effects and $\varepsilon_r$ is the observational/residual error. The matrices $\mathbf{X}$ and $\mathbf{Z}$ are incidence matrices that relate the various effects to observations. The first moments for the random effects (their expectations) are $E[\mathbf{u}] = 0$ and $E[\varepsilon_r] = 0$, and the variance-covariance structure is given by $\mathrm{var}[\mathbf{u}] = \mathbf{G}$, $\mathrm{var}[\varepsilon_r] = \mathbf{R}$ and $\mathrm{cov}[\mathbf{u}, \varepsilon_r] = 0$. The linear model can be introduced on a lattice in which case both Eq. (2) and Eq. (4) come into play simultaneously. In this case

$$\mathbf{G} = \sigma^2 (\mathbf{I} - \Delta)^{-1} [(\mathbf{I} - \Delta)^{-1}]'$$
$$\mathbf{R} = \sigma_1^2 \mathbf{I}$$

Additional assumptions are needed to implement maximum likelihood or computer simulation, and generally $\mathbf{y}$, $\mathbf{u}$, and $\varepsilon_r$ are taken as multivariate normal. As indicated in Goldberger (Ref. [5]), the Best Linear Unbiased Prediction (BLUP) of u is found by evaluating

$$\hat{\mathbf{u}} = \mathbf{G}\mathbf{Z}'\mathbf{V}^{-1}[\mathbf{y} - \mathbf{X}\hat{\beta}],$$

where

$$\mathbf{V} = \mathrm{var}(\mathbf{y}) = \mathbf{Z}\mathbf{G}\mathbf{Z}' + \mathbf{R} \qquad (5)$$

and where $\hat{\beta}$ is the Best Linear Unbiased Estimate (BLUE) of the fixed effects obtained by the Generalized Least Squares (GSE) problem

$$(\mathbf{X}'\mathbf{V}^{-1}\mathbf{X})[\hat{\beta}] = [\mathbf{X}'\mathbf{V}^{-1}\mathbf{y}]. \qquad (6)$$

These equations can be reformulated so that the solutions can be obtained directly from the mixed model equations (Ref. [6])

$$\begin{pmatrix} \mathbf{X}'\mathbf{R}^{-1}\mathbf{X} & \mathbf{X}'\mathbf{R}^{-1}\mathbf{Z} \\ \mathbf{Z}'\mathbf{R}^{-1}\mathbf{X} & \mathbf{Z}'\mathbf{R}^{-1}\mathbf{Z} + \mathbf{G}^{-1} \end{pmatrix} \begin{pmatrix} \hat{\beta} \\ \hat{\mathbf{u}} \end{pmatrix} = \begin{pmatrix} \mathbf{X}'\mathbf{R}^{-1}\mathbf{y} \\ \mathbf{Z}'\mathbf{R}^{-1}\mathbf{y} \end{pmatrix}$$

Associated with the mixed model equations of Eq. (8) is the Mixed Model Matrix, M,

$$\mathbf{M} = \begin{pmatrix} \mathbf{X}'\mathbf{R}^{-1}\mathbf{X} & \mathbf{X}'\mathbf{R}^{-1}\mathbf{Z} & \mathbf{X}'\mathbf{R}^{-1}\mathbf{y} \\ \mathbf{Z}'\mathbf{R}^{-1}\mathbf{X} & \mathbf{Z}'\mathbf{R}^{-1}\mathbf{Z} + \mathbf{G}^{-1} & \mathbf{Z}'\mathbf{R}^{-1}\mathbf{y} \\ \mathbf{y}'\mathbf{R}^{-1}\mathbf{X} & \mathbf{y}'\mathbf{R}^{-1}\mathbf{Z} & \mathbf{y}\mathbf{R}^{-1}\mathbf{y} \end{pmatrix} \qquad (8)$$

The log-likelihood for the Multivariate Normal (**MN**) applied to the lattice model is given by (see Ref. [15])

$$\ln(\mathbf{MN}) \propto -\frac{1}{2}\ln|\mathbf{V}| - \frac{1}{2}(\mathbf{y} - \mathbf{X}\beta)'\mathbf{V}^{-1}(\mathbf{y} - \mathbf{X}\beta) \qquad (9)$$

The maximum likelihood estimates of $\beta$ and the dispersion parameters ($\mathbf{R}$ and $\mathbf{G}$) are found by maximizing the log-likelihood. Estimates of the dispersion parameters can be badly biased by small-sample errors induced by the estimation of $\hat{\beta}$. This is a serious problem when the dimension of $\beta$ is large relative to the information available to estimate $\beta$.

To overcome this problem, Patterson and Thompson (Ref. [7]) introduced Restricted Maximum Likelihood (ReML), where the dispersion parameters are found

by maximizing

$$\ln(\mathbf{R}\,\mathrm{e}\,\mathbf{ML}) \propto -\frac{1}{2}\ln|\mathbf{R}| - \frac{1}{2}\ln|\mathbf{V}| - \frac{1}{2}\ln|\mathbf{X}'\mathbf{V}\mathbf{X}| - \frac{1}{2}(\mathbf{y}-\mathbf{X}\hat{\beta})'\mathbf{V}^{-1}(\mathbf{y}-\mathbf{X}\hat{\beta}) \qquad (10)$$

where $\hat{\beta}$ is the solution obtained by GSE, Eq. (6). ReML has the advantage of eliminating the $\beta$ parameters from the Likelihood. This is especially useful in cases where one wants to concentrate on minimizing the deviations from a common mean, without explicitly finding that common mean. Ref. [8] states, "In contrast to conventional maximum likelihood estimation, ReML can produce unbiased estimates of variance and covariance parameters." Harville (Ref. [9]) derived the likelihood in Eq. (11) to treat the "error contrasts" which are found by taking a complete set of linear combinations of the observations which are sufficient to remove the effect of $\beta$ while leaving the maximal amount of information for the purpose of ReML. An early review of ReML can be found in Ref. [10]. More recent reviews can be found in Ref. [11] and [12].

The above ReML likelihood can be put into an alternative form by applying a Bayesian analysis to Eq. (9) (see Ref. [15]) to obtain

$$\ln(\mathbf{ReML}) \propto -\frac{1}{2}\ln|\mathbf{R}| - \frac{1}{2}\ln|\mathbf{G}| - \frac{1}{2}\ln|\mathbf{C}| - \frac{1}{2}\mathbf{y}'\mathbf{V}^{-1}\mathbf{y}$$

In dealing with lattice models, it is often the case that $\mathbf{G}$ is a dense matrix, but $\mathbf{G}^{-1}$ is sparse. Therefore it is more expedient to work with Eq. (3) in terms of $\mathbf{G}^{-1}$, than with $\mathbf{G}$ directly. The term "$-\frac{1}{2}\ln|\mathbf{R}|$" is available by analytical methods because of the simple structure assumed for $\mathbf{R}$ ($\mathbf{R} = \sigma^2\mathbf{I}$ in most cases). The "$-\frac{1}{2}\ln|\mathbf{G}|$" term can be obtained by performing a Cholesky decomposition on $\mathbf{G}^{-1}$ followed by using the diagonal elements from the Cholesky decomposition to compute the determinant $|\mathbf{G}^{-1}|$ and subsequently using the relationship that $\mathbf{G} = 1/|\mathbf{G}^{-1}|$. Ref. [17] has shown that by performing Cholesky decomposition on M above, then the expression

$$-\frac{1}{2}\ln|\mathbf{C}| - \frac{1}{2}\mathbf{y}'\mathbf{V}_{\infty}^{-1}\mathbf{y}$$

can also be obtained.

Numerical methods exist for calculating the first and second derivatives of all these functions. The Hessian can be constructed and an iterative procedure involving the Newton-Raphson method can be used to maximize the likelihood.

It is well known that the Cholesky decomposition runs to completion with any matrix that is symmetric and non-negative definite. In the event of singular covariance structure, it is possible to generalize this method (Ref. [4]).

The likelihood function is derived from elements of the Cholesky decomposition, and so there is nothing else that is needed to perform ReML but to find the derivatives that permit optimization by the iterative Newton-Raphson technique applied to the derivatives of the ReML likelihood. These derivatives come automatically with the Cholesky decomposition (see Ref. [13], [14]) and so there is little beyond constructing the matrices $\mathbf{M}$ and $\mathbf{G}^{-1}$ and their Cholesky decompositions, that must be considered to describe ReML. The pseudo-code for the differentiation of the Cholesky algorithm and directions for how to use it are to be found in Ref. [17].

## 3. Results of fit and simulation with autocorrelations.

We used the above ReML method above to fit an empirical speckle pattern from data, for example, see Fig. 2. The ReML method was implemented on a lattice model using the following nearest-neighbor relationship in the intensity pattern, which is based on Eq. (2)

$$u_{ij} = \rho[u_{i+1,j} + u_{i-1,j} + u_{i,j+1} + u_{i,j-1}] + \varepsilon_{ij}. \qquad (12)$$

The first term on the Right-Hand Side of Eq. (12) contains the structure of $\mathbf{\Delta}$, which can be substituted into Eq. (3) in order to construct $\mathbf{G}^{-1}$. This $\mathbf{G}^{-1}$ together with $\mathbf{R}^{-1}$ can be used to construct the Mixed Model Matrix in Eq. (8) and the ReML likelihood function. This ReML likelihood can subsequently processed through the Newton-Raphson numerical method to find the best fit.

Our Gaussian spatial lattice fit to the data yielded $\rho = 0.10$ and a sample variance ratio of 0.454. Therefore there is a positive correlation in the pixel intensity pattern from nearest neighbor cells in qualitative agreement with the positive spatial Hurst parameter given in in Fig 4.

In order to correctly simulate such a pattern this positive correlation must be taken into account (see Ref. [18]).

With the fit done we were also able to use Monte Carlo techniques to simulate similar patterns including the correlated spatial structure as can be seen in Fig. 6 (the vertical scale is arbitrary based on the largest intensity).

The technique for simulation was based on simulating the diagonal elements with an N (0,1) Gaussian random number and using the Cholesky decomposition to solve the linear system for the original intensity variables. The nearest neighbor correlations are visible in the structure of the simulated distribution.

## 4. Conclusion

The method described in this the paper provides a way to analyze an actual speckle intensity pattern using a 2-dimensional spatial pattern on a lattice taking into account the nearest neighbor effects. The information can be summarized using the ReML method to estimate the parameters in the statistical model. The parameters characterize the statistical properties of the spatial speckle pattern, and estimating them permits statistical inferences having to do with variance and spatial correlations which are necessary to implement an accurate simulation (e.g., for SAR speckle simulation see Ref. 18). Completely new speckle patterns can be simulated from the spatial model that sets the parameters equal to the estimated parameters found from one data set.

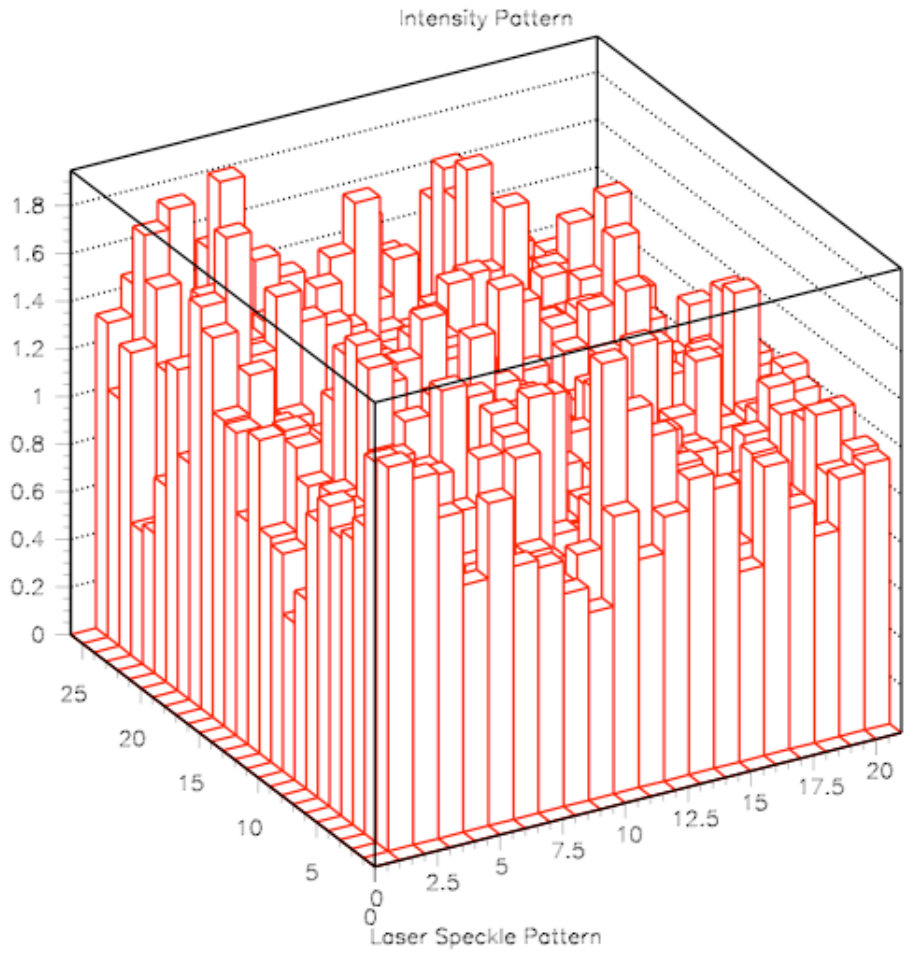

**Fig. 6.** Simulated speckle pattern in the intensity based on Gaussian model.